\documentclass{elsarticle}
\usepackage{color}
\usepackage{hyperref}

\begin{document}

\begin{frontmatter}

\title{What is the magnetic field distribution for the equation of state of magnetized neutron stars?}

\author[1]{V. Dexheimer}
\address[1]{Department of Physics, Kent State University, Kent OH 44242 USA}
\ead{vdexheim@kent.edu}

\author[2]{B. Franzon}
\address[2]{FIAS, Johann Wolfgang Goethe University, Frankfurt, Germany}

\author[3]{R. O. Gomes}
\address[3]{Instituto de F\'{\i}sica, Universidade Federal do Rio Grande do Sul, Porto Alegre, RS, Brazil}

\author[4]{R. L. S. Farias}
\address[4]{Departamento de F\'{\i}sica, Universidade Federal de Santa Maria, Santa Maria, RS, Brazil}

\author[5]{S. S. Avancini}
\address[5]{Departamento de F\'{\i}sica, Universidade Federal de Santa Catarina, Florian\'{o}polis, Brazil}

\author[2]{S. Schramm}

\begin{abstract}
In this Letter, we report a realistic calculation of the magnetic field profile for the equation of state inside strongly magnetized neutron stars. Unlike previous estimates, which are widely used in the literature, we find that magnetic fields increase relatively slowly with increasing baryon chemical potential (or baryon density) of magnetized matter. More precisely, the increase is polynomial instead of exponential, as previously assumed. Through the analysis of several different realistic models for the microscopic description of stellar matter (including hadronic, hybrid and quark models) combined with general relativistic solutions endowed with a poloidal magnetic field obtained by solving Einstein-Maxwell's field equations in a self-consistent way, we generate a phenomenological fit for the magnetic field distribution in the stellar polar direction to be used as input in microscopic calculations.
\end{abstract}

\begin{keyword}
\texttt Neutron Star \sep Equation of state \sep Quark deconfinement \sep Magnetic Field
\end{keyword}

\end{frontmatter}



In recent years, several measurements have shed new light on the strength of magnetic fields on the surface and in the interior of neutron stars. While measurements using anharmonic precession of star spin down have estimated surface magnetic fields to be on the magnitude of $10^{15}$ G for the sources 1E~1048.1-5937 and 1E~2259+586 \cite{Melatos:1999ji}, data for slow phase modulations in star hard x-ray pulsations (interpreted as free precession) suggest internal magnetic fields to be on the magnitude of $10^{16}$ G for the source 4U 0142+61 \cite{Makishima:2014dua}. Together, these estimates have motivated a large amount of research on the issue of how magnetic fields modify the microscopic structure (represented in the equation of state) and the macroscopic structure (obtained from the solution of Einstein-Maxwell's equations) of neutron stars.

In order to include the effect of magnetic fields in the equation of state to describe neutron stars, a profile for the strength of the field in a given direction has to be defined. Usually, this is done in two ways, both of which we will show to be incorrect. The first way is through the assumption of a constant magnetic field, which cannot be correct due to a simple magnetic field flux conservation assumption. The second, concerns assuming an ad hoc exponential formula for the field profile as a function of baryon density or baryon chemical potential. As already pointed out by Menezes  et al. in Ref.~\cite{Menezes:2016wbw}, ad hoc formulas for magnetic field profiles in neutron stars do not fulfill Maxwell's equations (more specifically, Gauss law) and, therefore, are incorrect. In this Letter, we present a realistic distribution for a poloidal magnetic field in the stellar polar direction as a function of a microscopic quantity, the baryon chemical potential. In order to do so, the macroscopic structure of the star obtained from the solution of Einstein-Maxwell equations has to be taken into account. In this way, we can ensure that the magnetic field distribution in the star respects the Einstein-Maxwell field equations.

In order to make our analysis as general as possible, in this work, we make use of three model equations of state for the microscopic description of neutron stars. They represent state-of-the-art approaches that include different assumptions of population in the core of  neutron stars, among other features. Two of them include magnetic field and anomalous magnetic moment effects, and we calculate (for each of these models) the equation of state as function of the magnetic field as an additional variable. In a second step, through the solution of Einstein's equations coupled with Maxwell's equations, we determine the magnetic field distribution in an individual star (with a fixed dipole magnetic moment), and then translate that to a field profile in the polar direction for the microscopic equation of state of each model. Later, we generalize one profile by averaging the results from the different models. All three models presented in the following fulfill current nuclear and astrophysical constraints, such as the prediction of massive stars. 

Note that, in this work, we are going to present results without the influence of temperature or star rotation. See Refs.~\cite{Franzon:2016iai,Negreiros:2011ak,Bejger:2016emu} and references therein for studies of the relation of magnetic field strengths, temperature  and rotation in the evolution of neutron stars. As has been discussed in Refs.~\cite{Dexheimer:2011pz,Sinha:2015bva,Tolos:2016hhl}, there is an important relation between magnetic field effects and star cooling, through the modification of the stellar population and the cooling processes themselves. The relations between rotation and magnetic fields in neutron stars have been studied in a general relativity approach in 
Refs.~\cite{Bocquet:1995je,Cardall:2000bs,Frieben:2012dz,Pili:2014npa}. In addition, it has been shown that toroidal fields are important for the stability of stellar magnetic fields \cite{prendergast1956equilibrium,markey1973adiabatic,tayler1973adiabatic,wright1973pinch,goldreich1992magnetic,Braithwaite:2005ps,Marchant:2010yj,Lasky:2011un,ciolfi2013twisted,Akgun:2013aq,mitchell2015instability,armaza2015magnetic,Mastrano:2015rfa}.
Nevertheless, in this case, the correspondence between magnetic field profiles in the equation of state and in the star is not straightforward, since magnetic fields are always included only in one direction in the microscopic description of stellar matter.


The first model we use was obtained from Refs.~\cite{Gomes:2014aka,Gomes:2014dka} by Gomes et al. and it will be referred to as ``G-model". It is a hadronic model that simulates many-body forces among nucleons by non-linear self-couplings and a field dependence on the interactions. The second model was obtained from Refs.~\cite{Dexheimer:2009hi,Dexheimer:2011pz} by Dexheimer et al. and it will be referred to as ``D-model". It includes nucleons, hyperons and quarks in a self-consistent approach and reproduces chiral symmetry restoration and deconfinement at high densities. The third model was obtained from Ref.~\cite{Hatsuda:1994pi} by Hatsuda et al. and it will be referred to as ``H-model". It is a version of the three-flavor NJL model that includes a repulsive vector-isoscalar interaction for the quarks, which is crucial for the description of astrophysical data (see Ref. \cite{Hanauske:2001nc} for an analysis of the repulsive quark interaction in neutron stars).

\begin{figure}[t]
\begin{centering}
\includegraphics[trim={1.4cm 0 0 0},width=9.7cm]{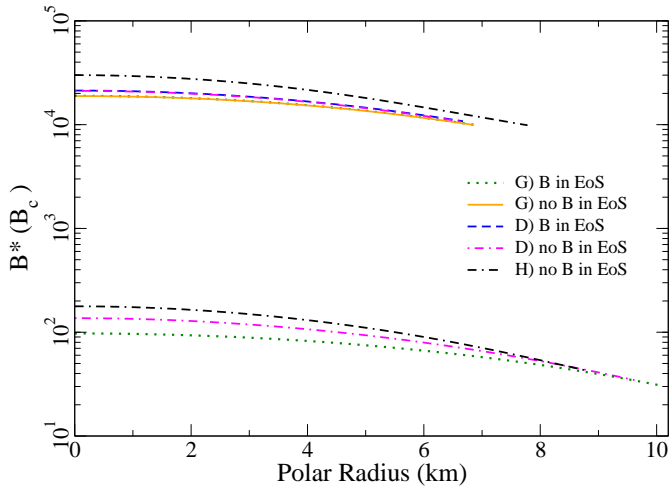}
\caption{(Color online) Magnetic field profile (in units of $B_c=4.414\times 10^{13}$ G) along the polar radius in a $M_B=2.2$ M$_\odot$ star obtained for the three equation of state models R, D and H. Each of these profiles is shown for a dipole magnetic moment $\mu=3\times10^{32}$ Am$^2$ (curves on the top) and for a dipole magnetic moment $\mu=1\times10^{30}$ Am$^2$ (curves on the bottom) including or not magnetic field effects in the equation of state. For the lower dipole magnetic moment, curves with and without effects in the equation of state completely overlap. \label{newfig1}}
\end{centering}
\end{figure}

The general-relativistic formalism used to describe the macroscopic features of magnetic neutron stars determines equilibrium configurations by solving the Einstein-Maxwell's field equations in spherical polar coordinates assuming a poloidal magnetic field configuration. For this purpose, we use the LORENE C++ class library for numerical relativity \cite{Bonazzola:1993zz,Bocquet:1995je,frieben2012equilibrium,Chatterjee:2014qsa,franzon2015self,Dexheimer:2017pom,rosana}. In this approach, the field is produced self-consistently by a macroscopic current, which is a function of the stellar radius, angle $\theta$ (with respect to symmetry axis), and dipole magnetic moment $\mu$ for each equation of state. The dipole magnetic moments shown in this work were chosen to reproduce a distribution with a central stellar magnetic field close to the upper limit of the code (maximum field strength that still reproduces a maximum density in the center of the star) and one to reproduce a surface magnetic field of about $10^{15}$ G, the maximum value observed on the surface of a star \cite{Melatos:1999ji}.

\begin{figure}[t!]
\begin{centering}
\includegraphics[trim={1.4cm 0 0 0},width=9.7cm]{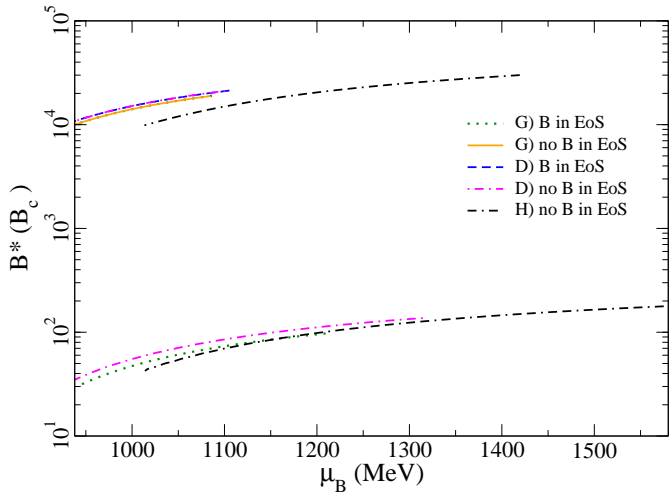}
\caption{(Color online) Magnetic field profile in the polar direction in a $M_B=2.2$ M$_\odot$ star as a function of baryon chemical potential obtained for the three equation of state models R, D and H. Each of these profiles is shown for a dipole magnetic moment $\mu=3\times10^{32}$ Am$^2$ (curves on the top) and for a dipole magnetic moment $\mu=1\times10^{30}$ Am$^2$ (curves on the bottom) including or not magnetic field effects in the equation of state. For the lower dipole magnetic moment, curves with and without effects in the equation of state completely overlap. \label{newfig3}}
\end{centering}
\end{figure}

\begin{table*}
\caption{\label{table1} Surface and central magnetic fields for the curves shown in the figures calculated for different baryonic mass stars, dipole magnetic moments and equations of state (without magnetic field effects in the equation of state, as they hardly change the field strength distribution).}
\begin{center}
\begin{tabular}{c|ccccc|ccccc}
$M_B$ (M$_\odot$)  & $\mu$ (Am$^2$)& EoS & $B_{surf}$ (G) & $B_{cent}$ (G)\\
\hline
$2.2$ & $3\times10^{32}$ & G & $4.49\times 10^{17}$ & $8.33\times 10^{17}$\\
$2.2$ & $3\times10^{32}$ & D & $4.73\times 10^{17}$ & $9.34\times 10^{17}$\\
$2.2$ & $3\times10^{32}$ & H & $4.14\times 10^{17}$ & $1.33\times 10^{18}$\\
\hline
$2.2$ & $1\times10^{30}$ & G & $1.34\times 10^{15}$ & $4.30\times 10^{15}$\\
$2.2$ & $1\times10^{30}$ & D & $1.53\times 10^{15}$ & $6.03\times 10^{15}$\\
$2.2$ & $1\times10^{30}$ & H & $1.87\times 10^{15}$ & $7.85\times 10^{15}$\\
\hline
$1.6$ & $2\times10^{32}$ & G & $2.84\times 10^{17} $ & $5.81\times 10^{17}$\\
$1.6$ & $2\times10^{32}$ & D & $2.87\times 10^{17}$ & $6.04\times 10^{17}$\\
$1.6$ & $2\times10^{32}$ & H & $1.03\times 10^{17}$ & $5.31\times 10^{17}$\\
\hline
$1.6$ & $1\times10^{30}$ & G & $1.34\times 10^{15}$ & $4.04\times 10^{15}$\\
$1.6$ & $1\times10^{30}$ & D & $1.24\times 10^{15}$ & $4.40\times 10^{15}$\\
$1.6$ & $1\times10^{30}$ & H & $4.84\times 10^{14}$ & $3.22\times 10^{15}$\\
\end{tabular}
\end{center}
\end{table*}

Figure~\ref{newfig1} shows the magnetic field profile obtained in the polar direction (in units of the critical field for the electron $B_c=4.414\times 10^{13}$ G) for the three equation of state models including the self-consistent solution of Einstein-Maxwell's equations. The values for the surface and central magnetic field strengths for the curves  are shown in Table.~\ref{table1}. The curves are shown for fixed values of the dipole magnetic moment $\mu$ including or not magnetic field effects in the equation of state for a star with baryonic mass $M_B=2.2$ M$_\odot$. It is important to note that, even when magnetic field effects are not included in the equation of state, the magnetic field still appears in the energy-momentum tensor through the magnetic energy, momentum density flux, and magnetic stress (see Eqs.~(4)-(8) of Ref.~\cite{Franzon:2015sya}). For the H-model, magnetic field effects could not be included in the equation of state due to the generation of a highly oscillating magnetization, as already pointed out in Refs.~\cite{Menezes:2009uc,Menezes:2015jka}.

There are two main conclusions that can be drawn from Figure~\ref{newfig1}. Firstly, whether or not one includes magnetic field effects in the equation of state of matter makes very little difference in the macroscopic magnetic field distribution of the star. It is important to note that magnetic field effects in the equation of state are still relevant for other quantities, such as the particle population and, consequently, the thermal evolution of neutron stars. Secondly, completely different equation of state models show different magnetic field strengths, but the respective profiles have approximately the same shape (when taking into account the logarithmic scale). The top curves of Figure~\ref{newfig1} are magnetic field profiles in the stellar polar direction for a higher dipole magnetic moment, while the bottom curves are profiles for a lower value of the dipole magnetic moment. In the latter case, whether or not one includes magnetic field effects in the equation of state makes no difference.

In Figure~\ref{newfig3}, we translate the magnetic field profile from Fig.~\ref{newfig1} into the thermodynamical quantity baryon chemical potential. The shape of the profiles obtained from the solution of Einstein-Maxwell's equations is well fit by a quadratic polynomial (and not exponential function). This allows us to fit one profile using the average of the different equation of state models. It depends only on the baryon chemical potential $\mu_B$ and on the value chosen for the dipole magnetic moment $\mu$
\begin{eqnarray}
B^*(\mu_B)=\frac{(a + b \mu_B + c \mu_B^2)}{B_c^2} \ \mu,
\label{3}
\end{eqnarray}
with coefficients $a$, $b$, and $c$ given in Table~\ref{table2}. In this case, $\mu_B$ should be given in MeV and $\mu$ in Am$^2$ in order to produce $B^*$ in units of the critical field for the electron $B_c=4.414\times 10^{13}$ G. Note that between Figs.~\ref{newfig1} and \ref{newfig3} the order of the EoS model curves change order (with respect to B*) as for model ``H" the same radius corresponds to much larger baryon chemical potentials. 

\begin{table*}
\caption{\label{table2} Quadratic fit coefficients $a$, $b$, and $c$ for Eq.~{1} calculated for different baryonic mass stars.}
\begin{center}
\begin{tabular}{c|ccc}
$M_B$ (M$_\odot$) \ \ & $a$ $\left(\frac{\rm{G}^2}{\rm{A m}^2}\right)$ & $b$ $\left(\frac{\rm{G}^2}{\rm{A m}^2 \rm{MeV}}\right)$ & $c$ $\left(\frac{\rm{G}^2}{\rm{A m}^2 \rm{MeV}^2}\right)$\\
\hline
$2.2$ & $-7.69\times10^{-1}$ & $1.20\times10^{-3}$ & $-3.46\times10^{-7}$\\
$1.6$ & $-1.02$ & $1.58\times10^{-3}$ & $-4.85\times10^{-7}$\\
\end{tabular}
\end{center}
\end{table*}

\begin{figure}[t!]
\begin{centering}
\includegraphics[trim={1.4cm 0 0 0},width=9.7cm]{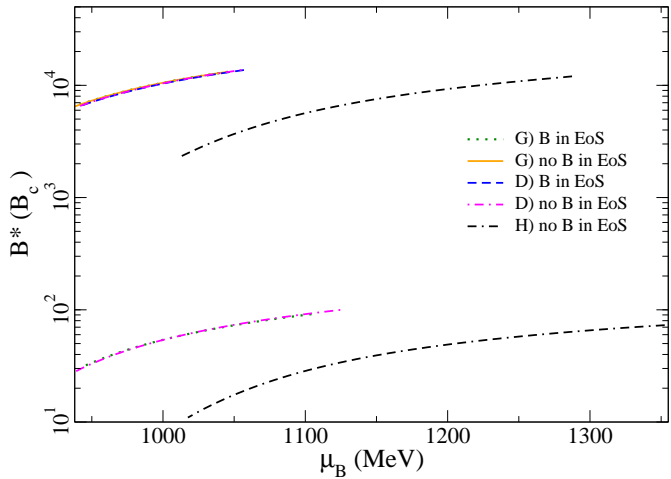}
\caption{(Color online) Same as Fig.~\ref{newfig3} but for a $M_B=1.6$ M$_\odot$ star with a dipole magnetic moment $\mu=2\times10^{32}$ Am$^2$ (curves on the top) and for a dipole magnetic moment $\mu=1\times10^{30}$ Am$^2$ (curves on the bottom) including or not magnetic field effects in the equation of state. Many of the curves overlap.\label{newfig4}}
\end{centering}
\end{figure}

But, what about other (lighter) stars? Each one would have about the same shape of magnetic field profile (again when taking into account the logarithmic scale) but with different strengths, as can be seen in Fig.~\ref{newfig4} for a $M_B=1.6$ M$_\odot$ star and different dipole magnetic moments. The values for the magnetic field strengths for the curves are shown in Table.~\ref{table1}. Note that a $M_B=1.6$ M$_\odot$ star is approximately equivalent to a canonical $M_G=1.4$ M$_\odot$ star. In this case, the parameters of the profile fit in Eq.~(\ref{3}) are once more given in Table~\ref{table2}, where it can be seen by the values of the parameter ``c" that the profiles for a larger star give on average a slightly more linear fit. Again, $\mu_B$ should be given in MeV and $\mu$ in Am$^2$ in order to produce $B^*$ in units of the critical field for the electron $B_c$. Note that, for a less massive star (and less compact), all equations of state that contain baryons reproduce very similar results. This stems from the fact that they were fitted to reproduce nuclear physics constraints and the central densities in such a star do not reach values much larger than saturation (less than two times $n_0$).

\begin{figure}[t!]
\begin{centering}
\includegraphics[trim={1.4cm 0 0 0},width=9.7cm]{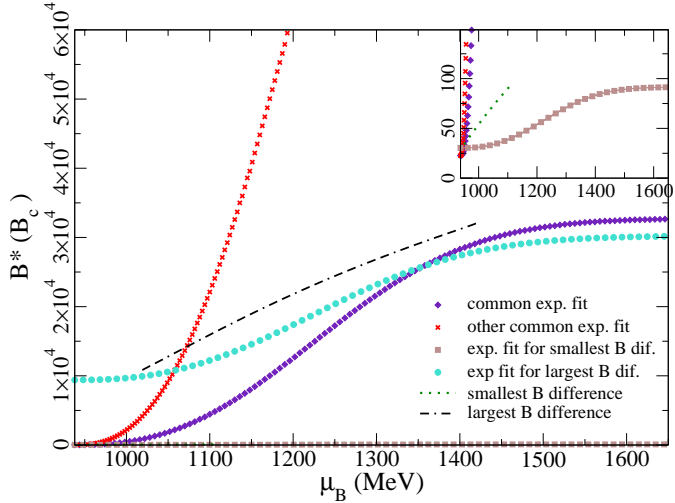}
\caption{(Color online) Comparison of our results with larger magnetic field strength variation (H with $M_B=2.2$ M$_\odot$ and $\mu=3\times10^{32}$ Am$^2$), smaller magnetic field strength variation (G with $M_B=1.6$ M$_\odot$ and $\mu=1\times10^{30}$ Am$^2$) and four ad hoc profiles from Eq.~\ref{eq1} (marked by symbols instead of lines). The ad hoc profiles were chosen to be two common ones (purple diamonds and red ``x"'s) and two that match on the surface and asymptotically the magnetic field strengths of our results (brown squares also shown in the inset and turquoise circles). The inset highlights the curves that reproduce lower magnetic field strengths. \label{newfig2}}
\end{centering}
\end{figure}

Only for comparison, we discuss now a figure (Fig.~\ref{newfig2}) containing the already mentioned ad hoc magnetic field strength exponential profiles. The original ansatz was written as a function of baryon density $n_B$
\begin{eqnarray}
B^*(n_B/n_0)=B_{surf}+B_0\left[1-e^{-\beta(n_B/n_0)^\gamma}\right],
\label{eq1}
\end{eqnarray}
with typical choices of constants $\beta=0.01$ and $\gamma =3$, surface magnetic field $B_{surf}$ and maximum field strength $B_{surf}+B_0$. This formula was suggested for the first time in Ref.~\cite{Bandyopadhyay:1997kh} but later rewritten as a function of baryon chemical potential (with the same structure) and subsequently used in about one hundred publications, among which the most cited ones are 
Refs.~\cite{Menezes:2008qt,Menezes:2009uc,Barkovich:2004jp,Dexheimer:2011pz,Rabhi:2009ih}.

Figure.~\ref{newfig2} shows a comparison of some of our results, the one reproducing the largest magnetic field strength variation (equation of state H for a star with $M_B=2.2$ M$_\odot$ and $\mu=3\times10^{32}$ Am$^2$), and the one reproducing the smallest magnetic field strength variation (equation of state G for a star with $M_B=1.6$ M$_\odot$ and $\mu=1\times10^{30}$ Am$^2$) together with four ad hoc profiles generated from Eq.~\ref{eq1}. (marked by symbols instead of lines). The ad hoc profiles were chosen to be two common ones (purple diamonds and red ``x"'s) and two that match on the surface and asymptotically the magnetic field strengths of our results (brown squares also shown in the inset and turquoise circles). The inset highlights the curves that reproduce lower magnetic field strengths. Clearly, none of the ad hoc exponential profiles coincide with our results (except maybe for one point), even the ad hoc profiles that were chosen to match our field strenghts on the surface of the star and at asymptotically high chemical potentials.

It is important to note that the fit we provide allows one to include a magnetic field profile in the polar direction in any equation of state in a simple way. This will allow analyses of magnetic field effects in specific models studying, for example,  changes in stiffness, changes in population, phase transitions, temperature (for fixed entropy per baryon), transport properties (thermal and electric conductivities), etc. A further inclusion of the obtained equations of state in a symmetric static isotropic solution for Einstein's equations (TOV \cite{Tolman:1939jz,Oppenheimer:1939ne}) to obtain macroscopic star properties is not a realistic approach when dealing with strong magnetic fields. This is because the magnetic field distribution is different and more complicated in other directions of the star and the pure magnetic field contribution would have to be added in an isotropic manner, being either positive or negative. In reality, this contribution has different signs in different directions and requires, therefore, a more advanced formalism (such as the one used in this Letter) which solves Einstein-Maxwell's field equations self-consistently.

At this point, one might be wondering why we did not choose to study profiles of magnetic fields with respect to baryon density. The reason is that the baryon density is not a continuous quantity in the presence of a first order phase transition (as a first derivative of the grand-potential) and, therefore, presents discontinuities. One exception concerns the construction of a mixed phase but, even in the case of an extended mixed phase in the star (like the one generated by the model referred to as ``D" in this work), a profile as a function of baryon density will present a change of slope at the mixed phase. The same argument can be applied to a magnetic field profile as a function of energy density, such as the one in Ref.~\cite{Lopes:2014vva}. In addition, different models present different relations between baryon chemical potential and baryon density or energy density, which makes it harder to construct one universal magnetic field fit function. Nevertheless, we note that, if we use a hadronic model such as the one referred to as ``G" in this work, the magnetic field profile calculated self-consistently as a function of baryon density will be a quartic polynomial but, still, not an exponential function.


In summary, we have provided a magnetic field profile as a function of baryon chemical potential (corresponding to the polar direction in a magnetized neutron star) which is to a large extent model independent. For this purpose, we have used three very different state-of-the-art equation of state models, built with different assumptions and including different degrees of freedom. When combined with the solutions of the Einstein-Maxwell's equations in a self-consistent way, this provided a formula to calculate how the magnetic field varies with baryon chemical potential, depending only on the dipole magnetic moment of choice and the stellar baryonic mass. A larger dipole magnetic moment produces a profile with larger magnetic field strengths for any baryon chemical potential. The resulting fit is quadratic in form and not exponential as previously assumed. Our fit is presented for the two most relevant types of neutron stars, one with gravitational mass around $2$ M$_\odot$ and one around $1.4$ M$_\odot$ (the baryonic mass correspondence varies slightly with model). The fit can be applied to any microscopic description of magnetized neutron stars. In this way, one can produce an equation of state that includes magnetic field effects without the need for solving the Einstein-Maxwell equations (as long as one is not interested in stellar macroscopic properties) but, still, not violating Gauss law.

In order to further refine our calculations, in the future, we intend to include temperature and rotation effects. In this way, we will be able to study the thermal evolution of magnetized proto-neutron and neutron stars. As already mentioned, toroidal magnetic field components are important for the long-term stability of magnetized neutron stars. Work on expanding the microscopic formalism to allow a more complex correspondence for magnetic fields in different directions of the star is underway.


The authors acknowledge support from NewCompStar COST Action MP1304 and from the LOEWE program HIC for FAIR. Work partially financed by CNPq under grants 308828/2013-5 (R.L.S.F) and 307458/2013-0 (S.S.A).

\section*{References}


\expandafter\ifx\csname url\endcsname\relax
  \def\url#1{\texttt{#1}}\fi
\expandafter\ifx\csname urlprefix\endcsname\relax\def\urlprefix{URL }\fi
\expandafter\ifx\csname href\endcsname\relax
  \def\href#1#2{#2} \def\path#1{#1}\fi


\bibliography{paper}


\end{document}